\documentclass[reprint,aps,pre,amsmath,amssymb,showpacs,nofootinbib,superscriptaddress]{revtex4-1}
\usepackage[pdftex]{graphicx} 
\usepackage{eucal}		
\usepackage{mycommands}

\newcommand{\rhostar}{\rho_*}

\newcommand{\cref}{c_\text{ref}}
\newcommand{\qhat}{\hat{q}}
\newcommand{\qhatturb}{\hat{q}^t}
\newcommand{\Tango}{\texttt{Tango}}

\providecommand{\eqnref}{}
\renewcommand{\eqnref}[1]{Eq.~\eqref{#1}}					

\begin{document}

\title{Bringing global gyrokinetic turbulence simulations to the transport timescale using a multiscale approach}

\author{Jeffrey B.~Parker}
\email{parker68@llnl.gov}
\affiliation{Lawrence Livermore National Laboratory, Livermore, CA 94550, USA}
\author{Lynda L.~LoDestro}
\affiliation{Lawrence Livermore National Laboratory, Livermore, CA 94550, USA}
\author{Daniel Told}
\affiliation{Max-Planck-Institut f\"{u}r Plasmaphysik, Boltzmannstra{\ss}e 2, 85748 Garching, Germany}
\author{Gabriele Merlo}
\affiliation{University of California, Los Angeles, California 90095, USA}
\author{Lee F.~Ricketson}
\affiliation{Lawrence Livermore National Laboratory, Livermore, CA 94550, USA}
\author{Alejandro Campos}
\affiliation{Lawrence Livermore National Laboratory, Livermore, CA 94550, USA}

\author{Frank Jenko}
\affiliation{Max-Planck-Institut f\"{u}r Plasmaphysik, Boltzmannstra{\ss}e 2, 85748 Garching, Germany}
\affiliation{University of California, Los Angeles, California 90095, USA}
\author{Jeffrey A.~F.~Hittinger}
\affiliation{Lawrence Livermore National Laboratory, Livermore, CA 94550, USA}


\begin{abstract}
The vast separation dividing the characteristic times of energy confinement and turbulence in the core of toroidal plasmas makes first-principles prediction on long timescales extremely challenging.  Here we report the demonstration of a multiple-timescale method that enables coupling global gyrokinetic simulations with a transport solver to calculate the evolution of the self-consistent temperature profile.  This method, which exhibits resiliency to the intrinsic fluctuations arising in turbulence simulations, holds potential for integrating nonlocal gyrokinetic turbulence simulations into predictive, whole-device models.
\end{abstract}

\maketitle

\section{Introduction}
First-principles simulations of tokamaks will play an increasingly important role in the mission to understand and control magnetized fusion plasmas.  Validated whole-device models aid in interpreting experimental results and help guide the planning of experiments to come.  Predictive modeling is particularly important for reactor-scale burning-plasma machines, with their enormous cost, enormous stored energy, and constrained number of shots.  While reduced modeling will be used for mapping out parameter space and initial shot planning in these devices, high-fidelity simulations will be required before finalizing shots, to ensure as well as possible that each shot will evolve as planned, and in particular that dangerous operating limits will not be approached.

One area of importance is high-fidelity prediction of heat and particle transport, which tend to be dominated by turbulent motions.  In this area, the workhorse framework is the gyrokinetic theory \cite{krommes:2012,brizard:2007}.  Gyrokinetic simulations of turbulence for tokamak core plasma have reached a state of maturity \cite{garbet:2010,jenko:2000,dorland:2000,candy:2003,chen:2007,grandgirard:2007,chang:2009,lin:2004,mcmillan:2010,gorler:2011b}, enabling a recent increase in emphasis on validating simulations to experiment \cite{rhodes:2011,jenko:2013,holland:2013,gorler:2014,citrin:2014,banon:2015,ernst:2016,howard:2016}.  These validation efforts have had some success; for example, when experimentally measured quantities such as the temperature and density profiles are provided as simulation inputs, numerical predictions of heat transport are often in quantitative agreement with experimentally inferred levels.  These results suggest that turbulence simulations are nearing a point where quantitative predictive capability is becoming plausible.

For such predictive value to fully materialize, gyrokinetic simulations must be incorporated into modeling on long experimental timescales.  Recently, some full-$f$ gyrokinetic simulations have been performed for an energy confinement time \cite{idomura:2014,idomura:2014b,difpradalier:2015,kim:2017a,kim:2017b}.  Because simulating to the energy confinement timescale is computationally expensive, however, gyrokinetic simulations often last for shorter durations and do not take into account the self-consistent evolution of plasma parameters occurring on the longer timescale.  Due to the difference in timescales between turbulence and transport, a direct approach to self-consistent simulation from turbulence to energy-confinement timescales is exceedingly challenging.  An estimate of the required computational resources, assuming gyro-Bohm scaling of the turbulent flux \cite{waltz:2005}, the use of magnetic-field-aligned coordinates, and a fixed radial resolution relative to the gyroradius, yields a computational-cost scaling as $\sim \rhostar^{-3}$, where $\rhostar$ is the ratio of the ion thermal gyroradius to the minor radius $a$ of the tokamak.  This cost renders such direct simulations of large reactors infeasible at the present time.  To illustrate, proceeding from transport-timescale gyrokinetic simulations of relatively small tokamaks ($\rhostar \approx 1/100$) to ITER ($\rhostar \approx 1/1000$)  \cite{shimada:2007} increases the computational cost of an already expensive scheme by three orders of magnitude.

A multiple-timescale approach could provide significantly more advantageous scaling to larger tokamaks such as ITER, by leapfrogging the timescale gap and reducing the computational-cost scaling from $\rhostar^{-3}$ to $\rhostar^{-1}$.  The $\rhostar^{-2}$ difference in cost scaling results from simulating turbulence phenomena and transport phenomena only on their natural timescales and avoiding the simulation of all intermediate timescales.  A multiple-timecale scale approach has previously been demonstrated with local, flux-tube gyrokinetic simulations \cite{candy:2009,barnes:2010}.  However, local simulations miss significant nonlocal effects such as turbulence spreading, transport avalanches, and transport in the core of high-performance discharges or internal transport barriers \cite{lin:2004,mcmillan:2010,gorler:2011b,ida:2015}.  A fully predictive capability requires global simulations.  Some early work focused on global gyrokinetic simulations with profile adjustments for matching experimental power balance \cite{waltz:2005b, waltz:2011}.  In this letter, we focus on global gyrokinetic simulations, presenting a unique method to bridge the timescale separation.

\section{Theoretical framework}
Beginning from the plasma Fokker--Planck equation and assuming a separation of timescales, one can derive coupled equations describing the fast microturbulence and the slowly evolving background Maxwellian.  Sugama and Horton first carried out this procedure, and more recently, Abel \emph{et al.}\ have elucidated it in greater detail \cite{sugama:1997,sugama:1998,abel:2013}.  We provide a brief overview here.

The formal orderings assume $\t_\text{turb}/ \t \sim \e^2$, $1/\W\t_\text{turb} \sim \e$, $\r/L \sim \rhostar \sim \e$, and $\de f/f_0 \sim \e$, where $\t_\text{turb}$ and $\t$ are the characteristic timescales of turbulence and transport, respectively, $\W$ is the ion cyclotron frequency, $\r$ is the ion gyroradius, $L$ is a characteristic scale length of the background density, temperature, or magnetic field, and $f_0$ and $\de f$ are the background and fluctuating distribution functions.  The smallness of $\e$ is well established in the core of a tokamak.  In the pedestal and edge region, where background profiles can vary on scales comparable to the gyroradius and fluctuations may be large, these orderings may break down.  An asymptotic expansion in $\e$ yields the gyrokinetic equation and evolution equations for the density, angular momentum, and pressure of each species.  The latter determine the density, rotation velocity, and temperature of the background Maxwellian $f_0$, which is in local thermal equilibrium on each flux surface.

On the microturbulence timescale, $f_0$ remains approximately constant.  The microturbulence and turbulent fluxes are assumed to be determined by the macroscopic profiles.  While the turbulence rapidly fluctuates, appropriate spatial and temporal averages over these quantities are assumed to evolve only on the slow timescale.

\subsection{Multiple-timescale numerical method}
Although gyrokinetic codes for simulating turbulence on the fast timescale are well established, solving the set of coupled gyrokinetic and transport equations is still an open problem.  Even though the gyrokinetic equation is five-dimensional (5D) and the transport equations are 1D, solving the transport equations self-consistently is quite challenging.  The difficulty is that the turbulent flux is strongly nonlinearly dependent on the profiles.  One is forced to confront this nonlinearity when using implicit time-stepping, which is highly desirable to enable long time steps.

We use the LoDestro method \cite{crotinger:1997,shestakov:2003} for coupling global gyrokinetic simulations to a transport equation.  We describe the method, which comprises several elements, as applied to the ion pressure equation:
	\begin{equation}
		\label{transportequation}
		\frac{3}{2} V' \pd{p}{t} + \pd{}{x} V' \qhat = V' S,
	\end{equation}
where $p$ is the ion pressure, $x$ is a radial flux coordinate, $V' = dV/dx$ is the differential volume of a flux surface, $\qhat = \avg{\v{Q} \cdot \nabla x}$ is the averaged radial projection of the heat flux, $\avg{\cdot}$ denotes a flux surface and intermediate time average, $\v{Q} = \int d\v{v}\, \tfrac{1}{2} mv^2 \v{v} f$ is the heat flux, and $S$ represents other local terms and sources.  For simplicity in this work, some of the terms in the full equation, such as turbulent heating and evolution of the magnetic equilibrium, have been neglected.  The heat flux includes both turbulent and neoclassical physics.

In the LoDestro method, the first element is to represent the turbulent component of $\qhat$, denoted $\qhatturb[p]$ to emphasize it is a functional of the pressure profile $p$, as the sum of diffusive and convective contributions.  Introducing a subscript $m$ to represent the time index and a subscript $l$ to represent the index of iteration within a timestep, we write
	\begin{equation}
		\qhatturb_{m,l} \to -D_{m,l-1} (\partial_x p_{m,l}) + c_{m,l-1} p_{m,l}.
		\label{fluxsplit}
	\end{equation}
This key step of lagging the effective transport coefficients $D$, $c$ (evaluating them at the previous iterate $l-1$) provides a practical scheme to iterate and converge to the nonlinear solution.  There is some freedom in determining how to split the flux between $D$ and $c$.  For instance, we write
	\begin{subequations}
	\label{fluxrepresentation}
	\begin{align}
		D_{m,l-1} &= -\th_{l-1} \qhatturb[p_{m,l-1}] \big/ \partial_x p_{m,l-1}, \\
		c_{m,l-1} &= (1-\th_{l-1}) \qhatturb[p_{m,l-1}] \big/ p_{m, l-1},
	\end{align}
	\end{subequations}
where one may choose the flux-split parameter $\th$ in a variety of ways.  Convective contributions can be important if the flux is nondiffusive.  Here, we choose $\th$ to be 1 or nearly 1, unless fluxes run up a gradient, in which case we choose $\th$ locally to be zero.  The iteration scheme becomes
	\begin{align}
		\label{iterationequation}
		\frac{3}{2} V' \frac{p_{m,l} - p_{m-1}}{\D t} + \partial_x \bigl[ &V' \bigl( -D_{m,l-1} \partial_x p_{m,l} \notag \\
			& + c_{m,l-1} p_{m,l} \bigr) \bigr] = V' S_m,
	\end{align}
where $\D t$ is the time step, and for notational convenience only the turbulent component of the flux has been written.  The scheme is fully time-dependent, although the first demonstration presented here seeks a steady-state.  Equation~\eref{iterationequation} is linear in the new iterate $p_{m,l}$ and hence straightforward to solve.  Despite the simplicity in representing the turbulent flux, if the iteration in $l$ converges, then a correct answer is obtained.  This can be seen by substituting \eqnref{fluxrepresentation} into \eqnref{fluxsplit}, in which case a converged answer means the representation for the flux on the right-hand side is equal to the actual turbulent flux on the left-hand side.

The second element modifies the basic idea in \eqnref{fluxrepresentation} in order to stabilize the iteration.  It was recognized in Ref.~\cite{shestakov:2003} that a diffusion coefficient's depending strongly on the gradient of the profile causes the iteration of Eqs.~(\ref{fluxrepresentation}--\ref{iterationequation}) to be numerically unstable.  The iteration can be stabilized by a relaxation of the diffusion coefficient, or similarly, of the turbulent flux.  Equation \eref{fluxrepresentation} is replaced by
	\begin{subequations}
	\label{fluxrepresentation_relaxation}
	\begin{align}
		D_{m,l-1} &= -\th_{l-1} \ol{\qhatturb}_{m,l-1} \big/ \partial_x \ol{p}_{m,l-1}, \\
		c_{m,l-1} &= (1-\th_{l-1}) \ol{\qhatturb}_{m,l-1} \big/ \ol{p}_{m, l-1},
	\end{align}
	\end{subequations}
where
	\begin{align}
		\ol{\qhatturb}_{m,l-1} &= \a \qhatturb [\ol{p}_{m, l-1}] + (1-\a) \ol{\qhatturb}_{m,l-2}, \label{relaxedflux} \\
		\ol{p}_{m,l-1} &= \a p_{m, l-1} + (1-\a)  \ol{p}_{m, l-2} \label{relaxedprofile}.
	\end{align}
In addition to stabilizing the iteration, a secondary purpose of the relaxation parameter $\a$ is to provide an effective averaging over previous iterates, reducing the statistical variance from the turbulence simulation.

The third element of the method attempts to minimize the amount of computational time spent in the turbulence simulation, which dominates the computational cost.  In our procedure, we run the turbulence simulation for a few eddy times and calculate an average flux $\qhatturb_{l-1}$ over that duration (suppressing the time-index subscript $m$ here for simplicity).  The relaxed flux $\ol{\qhatturb}_{l-1}$ is computed via \eqnref{relaxedflux}, which is then used to find a new profile iterate $p_l$ using Eqs.~(\ref{iterationequation}--\ref{fluxrepresentation_relaxation}) and a new $\ol{p}_l$ using \eqnref{relaxedprofile}.  Then, the process repeats for the next iteration.  The turbulence simulation runs with the new background profile $\ol{p}_l$ (background Maxwellian $f_0$), while using the turbulent state $\de f$ from the end of the previous iteration as the initial condition.

This procedure efficiently takes advantage of the $\de f$ representation in gyrokinetic simulations and avoids the need to start an entirely new simulation for each iteration.  The time-saving action of restarting a simulation with a new $f_0$ and the old $\de f$ can result in transient bursts associated with the fast relaxation to a self-consistent $\de f$, but if the variation in $f_0$ per iteration is small enough, the bursts have minor impact.  Computational efficiency is also promoted by averaging the flux over just a few turbulence eddy times per iteration.  This strategy may result in the turbulent flux lagging a couple iterations behind the profile iterates, but should not cause significant delay in convergence.

We now compare the LoDestro method with a Newton-based method, which was used by earlier works with \emph{local} gyrokinetic simulations \cite{candy:2009,barnes:2010}.  Those works developed an iteration scheme by expanding the flux as $\qhat_{l} = \qhat[p_l] \approx \qhat[p_{l-1}] + \de \qhat / \de p |_{p_{l-1}} \cdot   (p_l - p_{l-1})$.  Such a Newton-based approach, when applied to \emph{global} simulations, becomes much more challenging because the flux depends in principle on the profile over the entire radial domain, rather than just its local value and gradient as in the local limit.  The Jacobian $\de \qhat / \de p$ is therefore significantly higher dimensional; the  number of nonzero elements is proportional to $N_r^2$ in the global problem rather than to $N_r$ as in the local problem, where $N_r$ is the number of radial grid points.  Moreover, each Jacobian-vector product is an expensive prospect because it requires an additional full turbulence simulation to implement a finite-difference estimate of the derivative.  In addition to being costly, computation of a derivative by finite difference is challenging as it amplifies error in $\qhat$ arising from turbulent fluctuations.

The LoDestro method, which employs a variant of fixed-point, or Picard, iteration, avoids these pitfalls.  Not needing to project differences offers a large gain in efficiency, as the lower sensitivity to error allows the use of as little as a couple eddy times of gyrokinetic simulation per iteration.  This gain compensates for the primary disadvantage of fixed-point iteration, that an unaccelerated scheme theoretically achieves linear rather than superlinear asymptotic convergence. 
	
\subsection{Simulation Setup}
A newly developed 1D transport solver, \Tango, implements the LoDestro method.  \Tango, a Python code, is open source \cite{tangocode}.  Here, {\Tango} is coupled to the global gyrokinetic code GENE \cite{jenko:2000,gorler:2011,genecode} to simulate ion-temperature-gradient-driven turbulence.  

The GENE simulations are collisionless, electrostatic, and use deuterium ions and adiabatic electrons.  Circular magnetic geometry is used \cite{lapillonne:2009}, with safety factor $q = 0.868 + 2.2 \r^2$, where $\r = r/a$ and the radius $r$ plays the role of the flux coordinate.  The toroidal rotation velocity is set to zero.  The density profile is prescribed and fixed in time.  The electron temperature is held equal to the ion temperature.  GENE is run in the ``gradient-driven'' mode, in which a Krook-type source prevents the average profile from varying too much within an iteration as the gyrokinetic equation is integrated in time \cite{gorler:2011}.

{\Tango} solves the ion pressure equation in the domain $0 \le \r \le 0.9$, and GENE solves the gyrokinetic equation in the domain $0.1 \le \r \le 0.9$.  {\Tango} spatially discretizes with second-order finite differences and uses Neumann boundary conditions at the magnetic axis and Dirichlet conditions (fixed temperature) at the outer radius $\r = 0.9$.

Two simulations are described here, with key parameters given in Table \ref{tab:params}.  In each simulation, {\Tango} seeks a steady-state ($\D t = \infty$) solution to Eq.~\eref{transportequation}.  The tokamak in Simulation 1 is roughly half the size ($\rhostar \approx 1/149$) as that in Simulation 2 ($\rhostar \approx 1/292$).  In Simulation 1, $120 \times 16 \times 16 \times 78 \times 48$ grid points are used for the radial, binormal, coordinate along the magnetic field, parallel velocity, and magnetic moment dimensions, respectively, in GENE, and 27 radial grid points are used in \Tango.  In Simulation 2, $240 \times 16 \times 16 \times 78 \times 48$ grid points are used in GENE and 53 radial grid points are used in \Tango.  In lieu of a full neoclassical model, a small uniform diffusivity is included in the heat flux.  Further details on the simulations can be found in the Supplemental Material.


The setup for \Tango's simulation is as follows.  In each iteration, GENE runs for 50 $R_0 / v_{T_i}$, where $R_0$ is the major radius and $v_{T_i}$ is the ion thermal velocity.  The simulation is run for 50 iterations.  The relaxation parameter is $\a = 0.3$.  Prior to commencing iterations coupled with \Tango, GENE is run standalone for 600 $R_0 / v_{T_i}$ to seed the initial condition.


\begin{table}
\begin{center}
\caption{Key parameters for simulations.}
\label{tab:params}
\begin{tabular}{|lrr|}
\hline
Parameter & Simulation 1 & Simulation 2\\ \hline
Minor radius $a$ & 0.594 m & 1 m \\ 
Major radius $R_0$ & 1.65 m & 3 m \\
Input power $P_\text{in}$ & 3 MW & 20 MW \\
Added diffusivity & 0.05 m$^2$/s & 0.25 m$^2$/s \\
Outer $T$ boundary condition & 0.44 keV & 1 keV \\
$\rhostar$ & $1/149$ & 1/292 \\ \hline
\end{tabular}
\end{center}
\end{table}

\section{Results}
\begin{figure}[h]
		\centering
		\includegraphics{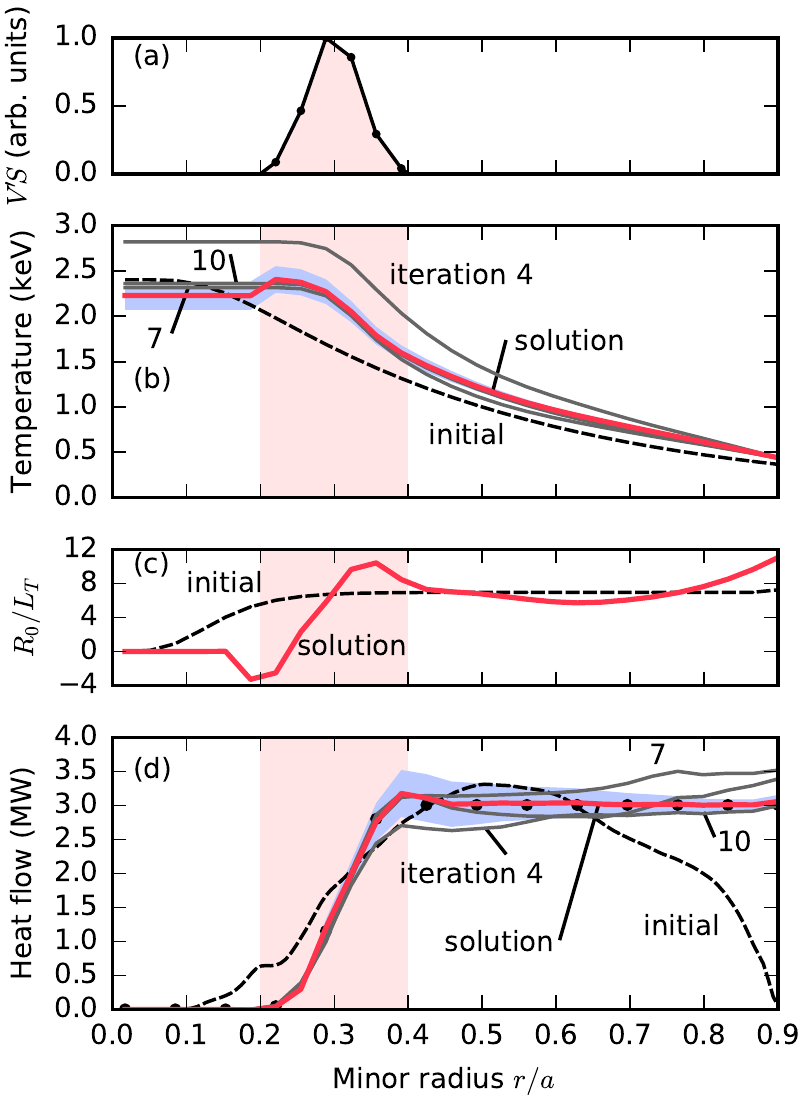}
		\caption{Results from Simulation 1.  (a) The 3 MW input heat source $V'S$.  The red shaded region indicates the area where the heat source is applied.  (b) Temperature profile.  (c) Normalized temperature gradient $R_0 / L_T = -(R_0/T) dT/dr$.  (d) Heat flow $H=V'\qhat$.  Shown in (b), (c), and (d) are the initial iterate (dashed curve), the mean over the final 20 out of 50 iterations (red curve labeled ``solution'').  Also shown in (b) and (d) are a few early iterations (numbered curves), and the standard deviation over the final 20 iterations (light blue shaded region).  In (d), the black dots depict the integral of the source, with which the numerical solution agrees well.  The initial iterate has a higher resolution than the other curves because it is taken directly from a GENE simulation rather than the {\Tango} grid.  This heat flow taken from GENE falls to zero at the outer edge because of the buffer zone in GENE.  The larger $R_0 / L_T$ that develops toward the edge $r/a \approx 0.9$ in Simulations 1 and 2 is an artifact of the buffer zone and does not have a direct effect (see the Supplemental Material for more details about the buffer zone).}
		\label{fig:smallmachine}
	\end{figure}

The results from Simulation 1 are shown in Figure~\ref{fig:smallmachine}, which depicts the applied heat source, the temperature profile, the normalized temperature gradient, and the total heat flow (turbulent $+$ added diffusivity) $H = V'\qhat$.  The initial condition (dashed curve) and a few early iterates (grey curves) are plotted for both temperature and heat flow.  After roughly 10--20 iterations, a noise floor is reached, where fluctuations in $\qhatturb$ prevent any further convergence.  More iterations do not substantially help because the relaxation in Eqs.~(\ref{relaxedflux}--\ref{relaxedprofile}) has a finite memory.

Focusing on iterates 31--50, we take the mean as the solution (red curve) and the standard deviation (light blue shaded region) as a rough measure of uncertainty.  The maximum coefficient of variation across radius is 7.1\% for the temperature.  For the heat flow, the maximum is 11.3\% for $\r \ge 0.28$; greater relative fluctuations occur at smaller $\r$ because the heat flow is near zero.  There is a small bump in temperature at $\r \approx 0.22$, at which the heat flux flows up the temperature gradient.  The bump is likely due to the strong localization of the heat source in this region.  The heat flux is still outward, an effect probably due to turbulence spreading from the adjacent region.
	
In steady state, the heat flow $H$ can be determined exactly by integrating Eq.~\eqref{transportequation} over radius; $H(r)$ must equal $\int_0^r d\ol{r}\, V'(\ol{r}) S(\ol{r})$.  The average of the numerically simulated heat flow over the final 20 iterations agrees well with this exact result (black dots).  The heat flows in both Simulation 1 and 2 are insensitive to the value of the ad-hoc diffusivity, as the heat flux resulting from this diffusivity is miniscule compared to the turbulent flux.
		
\begin{figure}[h]
		\centering
		\includegraphics{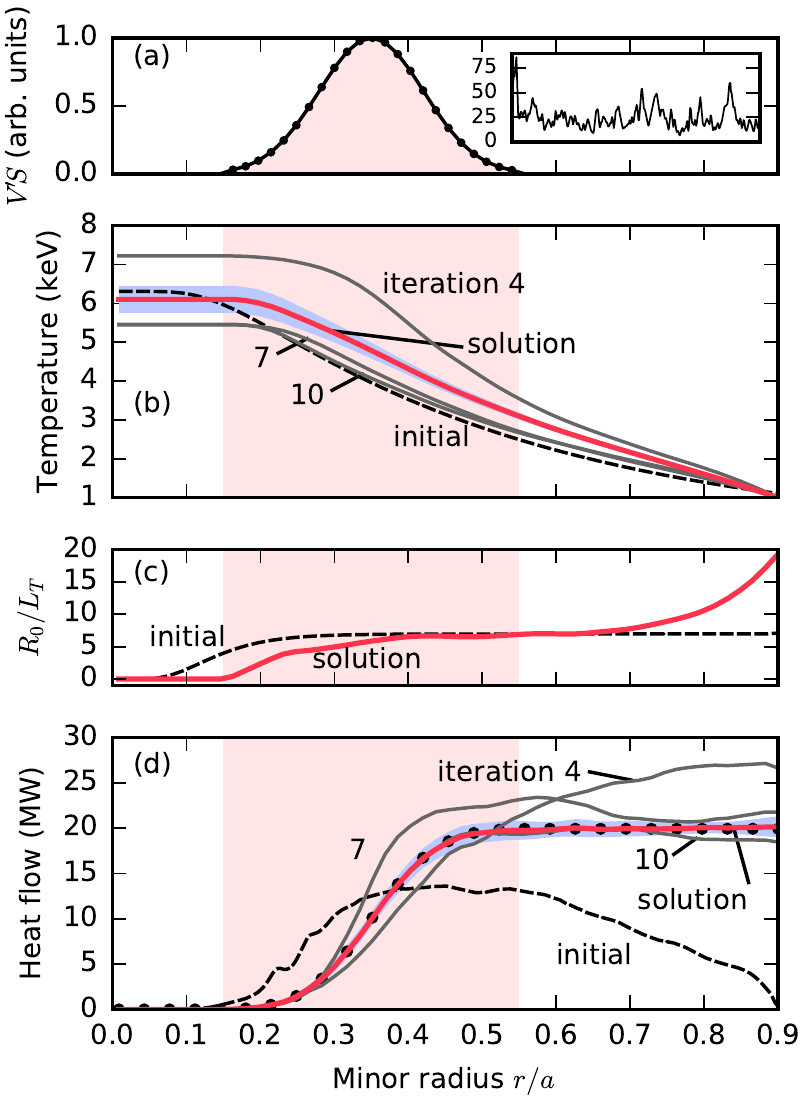}
		\caption{Results from Simulation 2, in the same format as Figure~\ref{fig:smallmachine}.  Here, 20 MW of input power is applied.  Inset: turbulent heat flux $\qhatturb$ (arb.\ units), centered at $r/a = 0.4$ and radially averaged over 10 gyroradii, shown as a function of time over 400 $R_0/v_{T_i}$ (simulated with the steady-state solution).}
		\label{fig:largemachine}
	\end{figure}

Figure~\ref{fig:largemachine} presents similar results for Simulation 2.  The multiscale coupling has also found the solution here, confirmed by the consistency of the numerical heat flow with the integral of the source, as shown in Figure~\ref{fig:largemachine}(d).  The maximum coefficient of variation across radius is 5.7\% for the temperature and 8\% for the heat flow ($\r \ge 0.35$).  It is interesting to note that the self-consistent heat flow is quite different from the initial heat flow, even though the final temperature profile does not appear that different from the initial temperature profile.  This behavior reflects the well-known nature of stiff transport, where small profile variations can lead to large changes in the flux.

\begin{figure}[h]
		\centering
		\includegraphics{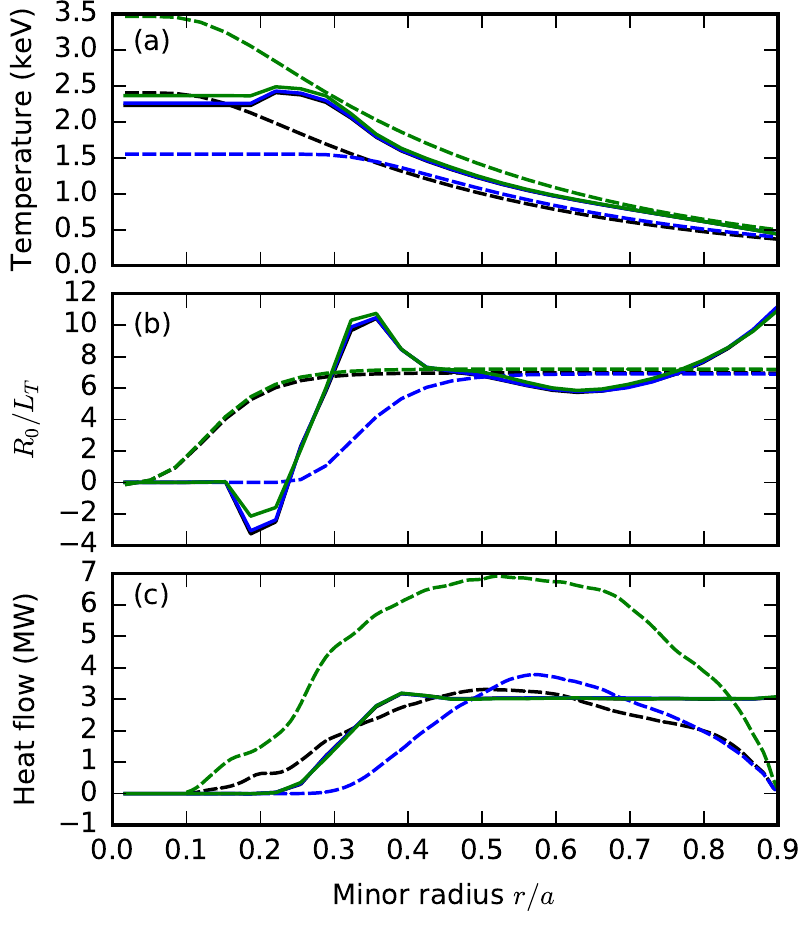}
		\caption{Comparing results when starting from different initial conditions, using the parameters of Simulation 1.  The initial temperature profile from Simulation 1 is used, along with two other distinct temperature profiles, distinguished by color.  Shown are the initial (dashed) and final, averaged over 20 iterations (solid) (a) temperature profile, (b) normalized temperature gradient, and (c) heat flow.  For the solutions, the solid curves are overlapping and are difficult to distinguish.}
		\label{fig:initialcondition}
	\end{figure}

Figure~\ref{fig:initialcondition} shows results of simulations, using the parameters of Simulation 1, but starting from three different initial temperature profiles.  The steady-state temperature profile, normalized temperature gradient, and heat flow obtained in the three cases are nearly identical.

As a test of sensitivity to numerical parameters, we varied the flux-split parameter $\th$ and redid Simulation 1 using constant $\th = 0.5$ or $\th=0.8$.  After iterating to convergence, we obtained the same steady-state solution as before.

\section{Discussion}
The energy confinement time can be estimated from $\t_E \approx W_i / P_\text{in}$, where $W_i$ is the stored energy in the ions in the steady-state solution.  We find $\t_E \approx 26$ ms for Simulation 1 and $\t_E \approx 55$ ms for Simulation 2.  The ratio of confinement times does not follow gyro-Bohm scaling $\t / \t_\text{turb} \sim \rhostar^{-2}$ exactly, as is known for small tokamaks \cite{candy:2004,mcmillan:2010}.

The total time used for gyrokinetic simulation is a useful measure of computational cost and efficiency of the multiscale method.  In both cases, this expenditure is 3100 $R_0 / v_{T_i}$, which corresponds to a physical time of 23 ms for both Simulations 1 and 2.  Hence, our calculation has successfully bridged the timescale separation.  Additionally, opportunity exists to optimize performance through parameter tuning.

Fluctuations in the turbulent flux create a tradeoff between computational expediency and precision.  An example of the fluctuations can be seen in Fig.~\ref{fig:largemachine}(a) (inset); the largest bursts can be three times as large as the mean.  These fluctuations, inherent to the equations regardless of the method of simulation, set a noise floor and limit the achievable precision.  The variance of the fluctuations can be reduced through greater averaging of the flux, which in practice is accomplished by increasing the simulation time per iteration or decreasing the relaxation parameter $\a$, either of which increases the computational cost.  It is worth noting that uncertainties in experimental measurements and limitations in physics models may set bounds on the precision desirable in practice, as additional precision requires greater computational resources while providing little added value beyond a certain point.

In summary, we have demonstrated multiple-timescale coupling of global gyrokinetic simulations to a transport solver.  Using the LoDestro method, which prescribes an algorithm for solving the nonlinear implicit timestep of the transport equation, we have solved for the steady-state temperature profile.  This method holds potential to play an important role in the integration of nonlocal gyrokinetic turbulence simulations into predictive, whole-device models.  Extensions of this work that allow for evolution of ion pressure, electron pressure, and density simultaneously are underway.

\begin{acknowledgments}
Useful discussions with Ian Abel are acknowledged.  This work was performed under the auspices of the U.S.\ Department of Energy by Lawrence Livermore National Laboratory under Contract No.\ DE-AC52-07NA27344, in part through the SciDAC Partnership for Multiscale Gyrokinetic Turbulence.  This research was supported by the Exascale Computing Project (17-SC-20-SC), a collaborative effort of the U.S. Department of Energy Office of Science and the National Nuclear Security Administration.  This research used resources of the National Energy Research Scientific Computing Center, a DOE Office of Science User Facility supported by the Office of Science of the U.S. Department of Energy under Contract No.\ DE-AC02-05CH11231.
\end{acknowledgments}

\bibliographystyle{apsrev4-1}
\bibliography{transport_turb} 

\begin{thebibliography}{42}%
\makeatletter
\providecommand \@ifxundefined [1]{%
 \@ifx{#1\undefined}
}%
\providecommand \@ifnum [1]{%
 \ifnum #1\expandafter \@firstoftwo
 \else \expandafter \@secondoftwo
 \fi
}%
\providecommand \@ifx [1]{%
 \ifx #1\expandafter \@firstoftwo
 \else \expandafter \@secondoftwo
 \fi
}%
\providecommand \natexlab [1]{#1}%
\providecommand \enquote  [1]{``#1''}%
\providecommand \bibnamefont  [1]{#1}%
\providecommand \bibfnamefont [1]{#1}%
\providecommand \citenamefont [1]{#1}%
\providecommand \href@noop [0]{\@secondoftwo}%
\providecommand \href [0]{\begingroup \@sanitize@url \@href}%
\providecommand \@href[1]{\@@startlink{#1}\@@href}%
\providecommand \@@href[1]{\endgroup#1\@@endlink}%
\providecommand \@sanitize@url [0]{\catcode `\\12\catcode `\$12\catcode
  `\&12\catcode `\#12\catcode `\^12\catcode `\_12\catcode `\%12\relax}%
\providecommand \@@startlink[1]{}%
\providecommand \@@endlink[0]{}%
\providecommand \url  [0]{\begingroup\@sanitize@url \@url }%
\providecommand \@url [1]{\endgroup\@href {#1}{\urlprefix }}%
\providecommand \urlprefix  [0]{URL }%
\providecommand \Eprint [0]{\href }%
\providecommand \doibase [0]{http://dx.doi.org/}%
\providecommand \selectlanguage [0]{\@gobble}%
\providecommand \bibinfo  [0]{\@secondoftwo}%
\providecommand \bibfield  [0]{\@secondoftwo}%
\providecommand \translation [1]{[#1]}%
\providecommand \BibitemOpen [0]{}%
\providecommand \bibitemStop [0]{}%
\providecommand \bibitemNoStop [0]{.\EOS\space}%
\providecommand \EOS [0]{\spacefactor3000\relax}%
\providecommand \BibitemShut  [1]{\csname bibitem#1\endcsname}%
\let\auto@bib@innerbib\@empty
\bibitem [{\citenamefont {Krommes}(2012)}]{krommes:2012}%
  \BibitemOpen
  \bibfield  {author} {\bibinfo {author} {\bibfnamefont {J.~A.}\ \bibnamefont
  {Krommes}},\ }\href@noop {} {\bibfield  {journal} {\bibinfo  {journal} {Annu.
  Rev. Fluid Mech.}\ }\textbf {\bibinfo {volume} {44}},\ \bibinfo {pages} {175}
  (\bibinfo {year} {2012})}\BibitemShut {NoStop}%
\bibitem [{\citenamefont {Brizard}\ and\ \citenamefont
  {Hahm}(2007)}]{brizard:2007}%
  \BibitemOpen
  \bibfield  {author} {\bibinfo {author} {\bibfnamefont {A.~J.}\ \bibnamefont
  {Brizard}}\ and\ \bibinfo {author} {\bibfnamefont {T.~S.}\ \bibnamefont
  {Hahm}},\ }\href {\doibase 10.1103/RevModPhys.79.421} {\bibfield  {journal}
  {\bibinfo  {journal} {Rev. Mod. Phys.}\ }\textbf {\bibinfo {volume} {79}},\
  \bibinfo {pages} {421} (\bibinfo {year} {2007})}\BibitemShut {NoStop}%
\bibitem [{\citenamefont {Garbet}\ \emph {et~al.}(2010)\citenamefont {Garbet},
  \citenamefont {Idomura}, \citenamefont {Villard},\ and\ \citenamefont
  {Watanabe}}]{garbet:2010}%
  \BibitemOpen
  \bibfield  {author} {\bibinfo {author} {\bibfnamefont {X.}~\bibnamefont
  {Garbet}}, \bibinfo {author} {\bibfnamefont {Y.}~\bibnamefont {Idomura}},
  \bibinfo {author} {\bibfnamefont {L.}~\bibnamefont {Villard}}, \ and\
  \bibinfo {author} {\bibfnamefont {T.}~\bibnamefont {Watanabe}},\ }\href
  {http://stacks.iop.org/0029-5515/50/i=4/a=043002} {\bibfield  {journal}
  {\bibinfo  {journal} {Nucl. Fusion}\ }\textbf {\bibinfo {volume} {50}},\
  \bibinfo {pages} {043002} (\bibinfo {year} {2010})}\BibitemShut {NoStop}%
\bibitem [{\citenamefont {Jenko}\ \emph {et~al.}(2000)\citenamefont {Jenko},
  \citenamefont {Dorland}, \citenamefont {Kotschenreuther},\ and\ \citenamefont
  {Rogers}}]{jenko:2000}%
  \BibitemOpen
  \bibfield  {author} {\bibinfo {author} {\bibfnamefont {F.}~\bibnamefont
  {Jenko}}, \bibinfo {author} {\bibfnamefont {W.}~\bibnamefont {Dorland}},
  \bibinfo {author} {\bibfnamefont {M.}~\bibnamefont {Kotschenreuther}}, \ and\
  \bibinfo {author} {\bibfnamefont {B.~N.}\ \bibnamefont {Rogers}},\ }\href
  {\doibase 10.1063/1.874014} {\bibfield  {journal} {\bibinfo  {journal} {Phys.
  Plasmas}\ }\textbf {\bibinfo {volume} {7}},\ \bibinfo {pages} {1904}
  (\bibinfo {year} {2000})}\BibitemShut {NoStop}%
\bibitem [{\citenamefont {Dorland}\ \emph {et~al.}(2000)\citenamefont
  {Dorland}, \citenamefont {Jenko}, \citenamefont {Kotschenreuther},\ and\
  \citenamefont {Rogers}}]{dorland:2000}%
  \BibitemOpen
  \bibfield  {author} {\bibinfo {author} {\bibfnamefont {W.}~\bibnamefont
  {Dorland}}, \bibinfo {author} {\bibfnamefont {F.}~\bibnamefont {Jenko}},
  \bibinfo {author} {\bibfnamefont {M.}~\bibnamefont {Kotschenreuther}}, \ and\
  \bibinfo {author} {\bibfnamefont {B.~N.}\ \bibnamefont {Rogers}},\ }\href
  {\doibase 10.1103/PhysRevLett.85.5579} {\bibfield  {journal} {\bibinfo
  {journal} {Phys. Rev. Lett.}\ }\textbf {\bibinfo {volume} {85}},\ \bibinfo
  {pages} {5579} (\bibinfo {year} {2000})}\BibitemShut {NoStop}%
\bibitem [{\citenamefont {Candy}\ and\ \citenamefont
  {Waltz}(2003)}]{candy:2003}%
  \BibitemOpen
  \bibfield  {author} {\bibinfo {author} {\bibfnamefont {J.}~\bibnamefont
  {Candy}}\ and\ \bibinfo {author} {\bibfnamefont {R.}~\bibnamefont {Waltz}},\
  }\href {\doibase http://dx.doi.org/10.1016/S0021-9991(03)00079-2} {\bibfield
  {journal} {\bibinfo  {journal} {J. Comput. Phys.}\ }\textbf {\bibinfo
  {volume} {186}},\ \bibinfo {pages} {545 } (\bibinfo {year}
  {2003})}\BibitemShut {NoStop}%
\bibitem [{\citenamefont {Chen}\ and\ \citenamefont
  {Parker}(2007)}]{chen:2007}%
  \BibitemOpen
  \bibfield  {author} {\bibinfo {author} {\bibfnamefont {Y.}~\bibnamefont
  {Chen}}\ and\ \bibinfo {author} {\bibfnamefont {S.~E.}\ \bibnamefont
  {Parker}},\ }\href {\doibase http://dx.doi.org/10.1016/j.jcp.2006.05.028}
  {\bibfield  {journal} {\bibinfo  {journal} {J. Comput. Phys.}\ }\textbf
  {\bibinfo {volume} {220}},\ \bibinfo {pages} {839 } (\bibinfo {year}
  {2007})}\BibitemShut {NoStop}%
\bibitem [{\citenamefont {Grandgirard}\ \emph {et~al.}(2007)\citenamefont
  {Grandgirard}, \citenamefont {Sarazin}, \citenamefont {Angelino},
  \citenamefont {Bottino}, \citenamefont {Crouseilles}, \citenamefont {Darmet},
  \citenamefont {Dif-Pradalier}, \citenamefont {Garbet}, \citenamefont
  {Ghendrih}, \citenamefont {Jolliet}, \citenamefont {Latu}, \citenamefont
  {Sonnendrücker},\ and\ \citenamefont {Villard}}]{grandgirard:2007}%
  \BibitemOpen
  \bibfield  {author} {\bibinfo {author} {\bibfnamefont {V.}~\bibnamefont
  {Grandgirard}}, \bibinfo {author} {\bibfnamefont {Y.}~\bibnamefont
  {Sarazin}}, \bibinfo {author} {\bibfnamefont {P.}~\bibnamefont {Angelino}},
  \bibinfo {author} {\bibfnamefont {A.}~\bibnamefont {Bottino}}, \bibinfo
  {author} {\bibfnamefont {N.}~\bibnamefont {Crouseilles}}, \bibinfo {author}
  {\bibfnamefont {G.}~\bibnamefont {Darmet}}, \bibinfo {author} {\bibfnamefont
  {G.}~\bibnamefont {Dif-Pradalier}}, \bibinfo {author} {\bibfnamefont
  {X.}~\bibnamefont {Garbet}}, \bibinfo {author} {\bibfnamefont
  {P.}~\bibnamefont {Ghendrih}}, \bibinfo {author} {\bibfnamefont
  {S.}~\bibnamefont {Jolliet}}, \bibinfo {author} {\bibfnamefont
  {G.}~\bibnamefont {Latu}}, \bibinfo {author} {\bibfnamefont {E.}~\bibnamefont
  {Sonnendrücker}}, \ and\ \bibinfo {author} {\bibfnamefont {L.}~\bibnamefont
  {Villard}},\ }\href {http://stacks.iop.org/0741-3335/49/i=12B/a=S16}
  {\bibfield  {journal} {\bibinfo  {journal} {Plasma Phys. Controlled Fusion}\
  }\textbf {\bibinfo {volume} {49}},\ \bibinfo {pages} {B173} (\bibinfo {year}
  {2007})}\BibitemShut {NoStop}%
\bibitem [{\citenamefont {Chang}\ \emph {et~al.}(2009)\citenamefont {Chang},
  \citenamefont {Ku}, \citenamefont {Diamond}, \citenamefont {Lin},
  \citenamefont {Parker}, \citenamefont {Hahm},\ and\ \citenamefont
  {Samatova}}]{chang:2009}%
  \BibitemOpen
  \bibfield  {author} {\bibinfo {author} {\bibfnamefont {C.~S.}\ \bibnamefont
  {Chang}}, \bibinfo {author} {\bibfnamefont {S.}~\bibnamefont {Ku}}, \bibinfo
  {author} {\bibfnamefont {P.~H.}\ \bibnamefont {Diamond}}, \bibinfo {author}
  {\bibfnamefont {Z.}~\bibnamefont {Lin}}, \bibinfo {author} {\bibfnamefont
  {S.}~\bibnamefont {Parker}}, \bibinfo {author} {\bibfnamefont {T.~S.}\
  \bibnamefont {Hahm}}, \ and\ \bibinfo {author} {\bibfnamefont
  {N.}~\bibnamefont {Samatova}},\ }\href {\doibase 10.1063/1.3099329}
  {\bibfield  {journal} {\bibinfo  {journal} {Phys. Plasmas}\ }\textbf
  {\bibinfo {volume} {16}},\ \bibinfo {pages} {056108} (\bibinfo {year}
  {2009})}\BibitemShut {NoStop}%
\bibitem [{\citenamefont {Lin}\ and\ \citenamefont {Hahm}(2004)}]{lin:2004}%
  \BibitemOpen
  \bibfield  {author} {\bibinfo {author} {\bibfnamefont {Z.}~\bibnamefont
  {Lin}}\ and\ \bibinfo {author} {\bibfnamefont {T.~S.}\ \bibnamefont {Hahm}},\
  }\href {\doibase 10.1063/1.1647136} {\bibfield  {journal} {\bibinfo
  {journal} {Phys. Plasmas}\ }\textbf {\bibinfo {volume} {11}},\ \bibinfo
  {pages} {1099} (\bibinfo {year} {2004})}\BibitemShut {NoStop}%
\bibitem [{\citenamefont {McMillan}\ \emph {et~al.}(2010)\citenamefont
  {McMillan}, \citenamefont {Lapillonne}, \citenamefont {Brunner},
  \citenamefont {Villard}, \citenamefont {Jolliet}, \citenamefont {Bottino},
  \citenamefont {G\"{o}rler},\ and\ \citenamefont {Jenko}}]{mcmillan:2010}%
  \BibitemOpen
  \bibfield  {author} {\bibinfo {author} {\bibfnamefont {B.~F.}\ \bibnamefont
  {McMillan}}, \bibinfo {author} {\bibfnamefont {X.}~\bibnamefont
  {Lapillonne}}, \bibinfo {author} {\bibfnamefont {S.}~\bibnamefont {Brunner}},
  \bibinfo {author} {\bibfnamefont {L.}~\bibnamefont {Villard}}, \bibinfo
  {author} {\bibfnamefont {S.}~\bibnamefont {Jolliet}}, \bibinfo {author}
  {\bibfnamefont {A.}~\bibnamefont {Bottino}}, \bibinfo {author} {\bibfnamefont
  {T.}~\bibnamefont {G\"{o}rler}}, \ and\ \bibinfo {author} {\bibfnamefont
  {F.}~\bibnamefont {Jenko}},\ }\href {\doibase 10.1103/PhysRevLett.105.155001}
  {\bibfield  {journal} {\bibinfo  {journal} {Phys. Rev. Lett.}\ }\textbf
  {\bibinfo {volume} {105}},\ \bibinfo {pages} {155001} (\bibinfo {year}
  {2010})}\BibitemShut {NoStop}%
\bibitem [{\citenamefont {G\"{o}rler}\ \emph
  {et~al.}(2011{\natexlab{a}})\citenamefont {G\"{o}rler}, \citenamefont
  {Lapillonne}, \citenamefont {Brunner}, \citenamefont {Dannert}, \citenamefont
  {Jenko}, \citenamefont {Aghdam}, \citenamefont {Marcus}, \citenamefont
  {McMillan}, \citenamefont {Merz}, \citenamefont {Sauter}, \citenamefont
  {Told},\ and\ \citenamefont {Villard}}]{gorler:2011b}%
  \BibitemOpen
  \bibfield  {author} {\bibinfo {author} {\bibfnamefont {T.}~\bibnamefont
  {G\"{o}rler}}, \bibinfo {author} {\bibfnamefont {X.}~\bibnamefont
  {Lapillonne}}, \bibinfo {author} {\bibfnamefont {S.}~\bibnamefont {Brunner}},
  \bibinfo {author} {\bibfnamefont {T.}~\bibnamefont {Dannert}}, \bibinfo
  {author} {\bibfnamefont {F.}~\bibnamefont {Jenko}}, \bibinfo {author}
  {\bibfnamefont {S.~K.}\ \bibnamefont {Aghdam}}, \bibinfo {author}
  {\bibfnamefont {P.}~\bibnamefont {Marcus}}, \bibinfo {author} {\bibfnamefont
  {B.~F.}\ \bibnamefont {McMillan}}, \bibinfo {author} {\bibfnamefont
  {F.}~\bibnamefont {Merz}}, \bibinfo {author} {\bibfnamefont {O.}~\bibnamefont
  {Sauter}}, \bibinfo {author} {\bibfnamefont {D.}~\bibnamefont {Told}}, \ and\
  \bibinfo {author} {\bibfnamefont {L.}~\bibnamefont {Villard}},\ }\href
  {\doibase 10.1063/1.3567484} {\bibfield  {journal} {\bibinfo  {journal}
  {Phys. Plasmas}\ }\textbf {\bibinfo {volume} {18}},\ \bibinfo {pages}
  {056103} (\bibinfo {year} {2011}{\natexlab{a}})}\BibitemShut {NoStop}%
\bibitem [{\citenamefont {Rhodes}\ \emph {et~al.}(2011)\citenamefont {Rhodes},
  \citenamefont {Holland}, \citenamefont {Smith}, \citenamefont {White},
  \citenamefont {Burrell}, \citenamefont {Candy}, \citenamefont {DeBoo},
  \citenamefont {Doyle}, \citenamefont {Hillesheim}, \citenamefont {Kinsey},
  \citenamefont {McKee}, \citenamefont {Mikkelsen}, \citenamefont {Peebles},
  \citenamefont {Petty}, \citenamefont {Prater}, \citenamefont {Parker},
  \citenamefont {Chen}, \citenamefont {Schmitz}, \citenamefont {Staebler},
  \citenamefont {Waltz}, \citenamefont {Wang}, \citenamefont {Yan},\ and\
  \citenamefont {Zeng}}]{rhodes:2011}%
  \BibitemOpen
  \bibfield  {author} {\bibinfo {author} {\bibfnamefont {T.}~\bibnamefont
  {Rhodes}}, \bibinfo {author} {\bibfnamefont {C.}~\bibnamefont {Holland}},
  \bibinfo {author} {\bibfnamefont {S.}~\bibnamefont {Smith}}, \bibinfo
  {author} {\bibfnamefont {A.}~\bibnamefont {White}}, \bibinfo {author}
  {\bibfnamefont {K.}~\bibnamefont {Burrell}}, \bibinfo {author} {\bibfnamefont
  {J.}~\bibnamefont {Candy}}, \bibinfo {author} {\bibfnamefont
  {J.}~\bibnamefont {DeBoo}}, \bibinfo {author} {\bibfnamefont
  {E.}~\bibnamefont {Doyle}}, \bibinfo {author} {\bibfnamefont
  {J.}~\bibnamefont {Hillesheim}}, \bibinfo {author} {\bibfnamefont
  {J.}~\bibnamefont {Kinsey}}, \bibinfo {author} {\bibfnamefont
  {G.}~\bibnamefont {McKee}}, \bibinfo {author} {\bibfnamefont
  {D.}~\bibnamefont {Mikkelsen}}, \bibinfo {author} {\bibfnamefont
  {W.}~\bibnamefont {Peebles}}, \bibinfo {author} {\bibfnamefont
  {C.}~\bibnamefont {Petty}}, \bibinfo {author} {\bibfnamefont
  {R.}~\bibnamefont {Prater}}, \bibinfo {author} {\bibfnamefont
  {S.}~\bibnamefont {Parker}}, \bibinfo {author} {\bibfnamefont
  {Y.}~\bibnamefont {Chen}}, \bibinfo {author} {\bibfnamefont {L.}~\bibnamefont
  {Schmitz}}, \bibinfo {author} {\bibfnamefont {G.}~\bibnamefont {Staebler}},
  \bibinfo {author} {\bibfnamefont {R.}~\bibnamefont {Waltz}}, \bibinfo
  {author} {\bibfnamefont {G.}~\bibnamefont {Wang}}, \bibinfo {author}
  {\bibfnamefont {Z.}~\bibnamefont {Yan}}, \ and\ \bibinfo {author}
  {\bibfnamefont {L.}~\bibnamefont {Zeng}},\ }\href
  {http://stacks.iop.org/0029-5515/51/i=6/a=063022} {\bibfield  {journal}
  {\bibinfo  {journal} {Nucl. Fusion}\ }\textbf {\bibinfo {volume} {51}},\
  \bibinfo {pages} {063022} (\bibinfo {year} {2011})}\BibitemShut {NoStop}%
\bibitem [{\citenamefont {Jenko}\ \emph {et~al.}(2013)\citenamefont {Jenko},
  \citenamefont {Told}, \citenamefont {G\"{o}rler}, \citenamefont {Citrin},
  \citenamefont {Navarro}, \citenamefont {Bourdelle}, \citenamefont {Brunner},
  \citenamefont {Conway}, \citenamefont {Dannert}, \citenamefont {Doerk},
  \citenamefont {Hatch}, \citenamefont {Haverkort}, \citenamefont {Hobirk},
  \citenamefont {Hogeweij}, \citenamefont {Mantica}, \citenamefont {Pueschel},
  \citenamefont {Sauter}, \citenamefont {Villard}, \citenamefont {Wolfrum},\
  and\ \citenamefont {the ASDEX Upgrade~Team}}]{jenko:2013}%
  \BibitemOpen
  \bibfield  {author} {\bibinfo {author} {\bibfnamefont {F.}~\bibnamefont
  {Jenko}}, \bibinfo {author} {\bibfnamefont {D.}~\bibnamefont {Told}},
  \bibinfo {author} {\bibfnamefont {T.}~\bibnamefont {G\"{o}rler}}, \bibinfo
  {author} {\bibfnamefont {J.}~\bibnamefont {Citrin}}, \bibinfo {author}
  {\bibfnamefont {A.~B.}\ \bibnamefont {Navarro}}, \bibinfo {author}
  {\bibfnamefont {C.}~\bibnamefont {Bourdelle}}, \bibinfo {author}
  {\bibfnamefont {S.}~\bibnamefont {Brunner}}, \bibinfo {author} {\bibfnamefont
  {G.}~\bibnamefont {Conway}}, \bibinfo {author} {\bibfnamefont
  {T.}~\bibnamefont {Dannert}}, \bibinfo {author} {\bibfnamefont
  {H.}~\bibnamefont {Doerk}}, \bibinfo {author} {\bibfnamefont
  {D.}~\bibnamefont {Hatch}}, \bibinfo {author} {\bibfnamefont
  {J.}~\bibnamefont {Haverkort}}, \bibinfo {author} {\bibfnamefont
  {J.}~\bibnamefont {Hobirk}}, \bibinfo {author} {\bibfnamefont
  {G.}~\bibnamefont {Hogeweij}}, \bibinfo {author} {\bibfnamefont
  {P.}~\bibnamefont {Mantica}}, \bibinfo {author} {\bibfnamefont
  {M.}~\bibnamefont {Pueschel}}, \bibinfo {author} {\bibfnamefont
  {O.}~\bibnamefont {Sauter}}, \bibinfo {author} {\bibfnamefont
  {L.}~\bibnamefont {Villard}}, \bibinfo {author} {\bibfnamefont
  {E.}~\bibnamefont {Wolfrum}}, \ and\ \bibinfo {author} {\bibnamefont {the
  ASDEX Upgrade~Team}},\ }\href
  {http://stacks.iop.org/0029-5515/53/i=7/a=073003} {\bibfield  {journal}
  {\bibinfo  {journal} {Nucl. Fusion}\ }\textbf {\bibinfo {volume} {53}},\
  \bibinfo {pages} {073003} (\bibinfo {year} {2013})}\BibitemShut {NoStop}%
\bibitem [{\citenamefont {Holland}\ \emph {et~al.}(2013)\citenamefont
  {Holland}, \citenamefont {Kinsey}, \citenamefont {DeBoo}, \citenamefont
  {Burrell}, \citenamefont {Luce}, \citenamefont {Smith}, \citenamefont
  {Petty}, \citenamefont {White}, \citenamefont {Rhodes}, \citenamefont
  {Schmitz}, \citenamefont {Doyle}, \citenamefont {Hillesheim}, \citenamefont
  {McKee}, \citenamefont {Yan}, \citenamefont {Wang}, \citenamefont {Zeng},
  \citenamefont {Grierson}, \citenamefont {Marinoni}, \citenamefont {Mantica},
  \citenamefont {Snyder}, \citenamefont {Waltz}, \citenamefont {Staebler},\
  and\ \citenamefont {Candy}}]{holland:2013}%
  \BibitemOpen
  \bibfield  {author} {\bibinfo {author} {\bibfnamefont {C.}~\bibnamefont
  {Holland}}, \bibinfo {author} {\bibfnamefont {J.}~\bibnamefont {Kinsey}},
  \bibinfo {author} {\bibfnamefont {J.}~\bibnamefont {DeBoo}}, \bibinfo
  {author} {\bibfnamefont {K.}~\bibnamefont {Burrell}}, \bibinfo {author}
  {\bibfnamefont {T.}~\bibnamefont {Luce}}, \bibinfo {author} {\bibfnamefont
  {S.}~\bibnamefont {Smith}}, \bibinfo {author} {\bibfnamefont
  {C.}~\bibnamefont {Petty}}, \bibinfo {author} {\bibfnamefont
  {A.}~\bibnamefont {White}}, \bibinfo {author} {\bibfnamefont
  {T.}~\bibnamefont {Rhodes}}, \bibinfo {author} {\bibfnamefont
  {L.}~\bibnamefont {Schmitz}}, \bibinfo {author} {\bibfnamefont
  {E.}~\bibnamefont {Doyle}}, \bibinfo {author} {\bibfnamefont
  {J.}~\bibnamefont {Hillesheim}}, \bibinfo {author} {\bibfnamefont
  {G.}~\bibnamefont {McKee}}, \bibinfo {author} {\bibfnamefont
  {Z.}~\bibnamefont {Yan}}, \bibinfo {author} {\bibfnamefont {G.}~\bibnamefont
  {Wang}}, \bibinfo {author} {\bibfnamefont {L.}~\bibnamefont {Zeng}}, \bibinfo
  {author} {\bibfnamefont {B.}~\bibnamefont {Grierson}}, \bibinfo {author}
  {\bibfnamefont {A.}~\bibnamefont {Marinoni}}, \bibinfo {author}
  {\bibfnamefont {P.}~\bibnamefont {Mantica}}, \bibinfo {author} {\bibfnamefont
  {P.}~\bibnamefont {Snyder}}, \bibinfo {author} {\bibfnamefont
  {R.}~\bibnamefont {Waltz}}, \bibinfo {author} {\bibfnamefont
  {G.}~\bibnamefont {Staebler}}, \ and\ \bibinfo {author} {\bibfnamefont
  {J.}~\bibnamefont {Candy}},\ }\href
  {http://stacks.iop.org/0029-5515/53/i=8/a=083027} {\bibfield  {journal}
  {\bibinfo  {journal} {Nucl. Fusion}\ }\textbf {\bibinfo {volume} {53}},\
  \bibinfo {pages} {083027} (\bibinfo {year} {2013})}\BibitemShut {NoStop}%
\bibitem [{\citenamefont {G\"{o}rler}\ \emph {et~al.}(2014)\citenamefont
  {G\"{o}rler}, \citenamefont {White}, \citenamefont {Told}, \citenamefont
  {Jenko}, \citenamefont {Holland},\ and\ \citenamefont
  {Rhodes}}]{gorler:2014}%
  \BibitemOpen
  \bibfield  {author} {\bibinfo {author} {\bibfnamefont {T.}~\bibnamefont
  {G\"{o}rler}}, \bibinfo {author} {\bibfnamefont {A.~E.}\ \bibnamefont
  {White}}, \bibinfo {author} {\bibfnamefont {D.}~\bibnamefont {Told}},
  \bibinfo {author} {\bibfnamefont {F.}~\bibnamefont {Jenko}}, \bibinfo
  {author} {\bibfnamefont {C.}~\bibnamefont {Holland}}, \ and\ \bibinfo
  {author} {\bibfnamefont {T.~L.}\ \bibnamefont {Rhodes}},\ }\href {\doibase
  10.1063/1.4904301} {\bibfield  {journal} {\bibinfo  {journal} {Phys.
  Plasmas}\ }\textbf {\bibinfo {volume} {21}},\ \bibinfo {pages} {122307}
  (\bibinfo {year} {2014})}\BibitemShut {NoStop}%
\bibitem [{\citenamefont {Citrin}\ \emph {et~al.}(2014)\citenamefont {Citrin},
  \citenamefont {Jenko}, \citenamefont {Mantica}, \citenamefont {Told},
  \citenamefont {Bourdelle}, \citenamefont {Dumont}, \citenamefont {Garcia},
  \citenamefont {Haverkort}, \citenamefont {Hogeweij}, \citenamefont {Johnson},
  \citenamefont {Pueschel},\ and\ \citenamefont {contributors}}]{citrin:2014}%
  \BibitemOpen
  \bibfield  {author} {\bibinfo {author} {\bibfnamefont {J.}~\bibnamefont
  {Citrin}}, \bibinfo {author} {\bibfnamefont {F.}~\bibnamefont {Jenko}},
  \bibinfo {author} {\bibfnamefont {P.}~\bibnamefont {Mantica}}, \bibinfo
  {author} {\bibfnamefont {D.}~\bibnamefont {Told}}, \bibinfo {author}
  {\bibfnamefont {C.}~\bibnamefont {Bourdelle}}, \bibinfo {author}
  {\bibfnamefont {R.}~\bibnamefont {Dumont}}, \bibinfo {author} {\bibfnamefont
  {J.}~\bibnamefont {Garcia}}, \bibinfo {author} {\bibfnamefont
  {J.}~\bibnamefont {Haverkort}}, \bibinfo {author} {\bibfnamefont
  {G.}~\bibnamefont {Hogeweij}}, \bibinfo {author} {\bibfnamefont
  {T.}~\bibnamefont {Johnson}}, \bibinfo {author} {\bibfnamefont
  {M.}~\bibnamefont {Pueschel}}, \ and\ \bibinfo {author} {\bibfnamefont
  {J.-E.}\ \bibnamefont {contributors}},\ }\href
  {http://stacks.iop.org/0029-5515/54/i=2/a=023008} {\bibfield  {journal}
  {\bibinfo  {journal} {Nucl. Fusion}\ }\textbf {\bibinfo {volume} {54}},\
  \bibinfo {pages} {023008} (\bibinfo {year} {2014})}\BibitemShut {NoStop}%
\bibitem [{\citenamefont {{Ba{\~n}{\'o}n Navarro}}\ \emph
  {et~al.}(2015)\citenamefont {{Ba{\~n}{\'o}n Navarro}}, \citenamefont
  {Happel}, \citenamefont {G\"{o}rler}, \citenamefont {Jenko}, \citenamefont
  {Abiteboul}, \citenamefont {Bustos}, \citenamefont {Doerk},\ and\
  \citenamefont {Told}}]{banon:2015}%
  \BibitemOpen
  \bibfield  {author} {\bibinfo {author} {\bibfnamefont {A.}~\bibnamefont
  {{Ba{\~n}{\'o}n Navarro}}}, \bibinfo {author} {\bibfnamefont
  {T.}~\bibnamefont {Happel}}, \bibinfo {author} {\bibfnamefont
  {T.}~\bibnamefont {G\"{o}rler}}, \bibinfo {author} {\bibfnamefont
  {F.}~\bibnamefont {Jenko}}, \bibinfo {author} {\bibfnamefont
  {J.}~\bibnamefont {Abiteboul}}, \bibinfo {author} {\bibfnamefont
  {A.}~\bibnamefont {Bustos}}, \bibinfo {author} {\bibfnamefont
  {H.}~\bibnamefont {Doerk}}, \ and\ \bibinfo {author} {\bibfnamefont
  {D.}~\bibnamefont {Told}},\ }\href {\doibase 10.1063/1.4919022} {\bibfield
  {journal} {\bibinfo  {journal} {Phys. Plasmas}\ }\textbf {\bibinfo {volume}
  {22}},\ \bibinfo {pages} {042513} (\bibinfo {year} {2015})}\BibitemShut
  {NoStop}%
\bibitem [{\citenamefont {Ernst}\ \emph {et~al.}(2016)\citenamefont {Ernst},
  \citenamefont {Burrell}, \citenamefont {Guttenfelder}, \citenamefont
  {Rhodes}, \citenamefont {Dimits}, \citenamefont {Bravenec}, \citenamefont
  {Grierson}, \citenamefont {Holland}, \citenamefont {Lohr}, \citenamefont
  {Marinoni}, \citenamefont {McKee}, \citenamefont {Petty}, \citenamefont
  {Rost}, \citenamefont {Schmitz}, \citenamefont {Wang}, \citenamefont
  {Zemedkun},\ and\ \citenamefont {Zeng}}]{ernst:2016}%
  \BibitemOpen
  \bibfield  {author} {\bibinfo {author} {\bibfnamefont {D.~R.}\ \bibnamefont
  {Ernst}}, \bibinfo {author} {\bibfnamefont {K.~H.}\ \bibnamefont {Burrell}},
  \bibinfo {author} {\bibfnamefont {W.}~\bibnamefont {Guttenfelder}}, \bibinfo
  {author} {\bibfnamefont {T.~L.}\ \bibnamefont {Rhodes}}, \bibinfo {author}
  {\bibfnamefont {A.~M.}\ \bibnamefont {Dimits}}, \bibinfo {author}
  {\bibfnamefont {R.}~\bibnamefont {Bravenec}}, \bibinfo {author}
  {\bibfnamefont {B.~A.}\ \bibnamefont {Grierson}}, \bibinfo {author}
  {\bibfnamefont {C.}~\bibnamefont {Holland}}, \bibinfo {author} {\bibfnamefont
  {J.}~\bibnamefont {Lohr}}, \bibinfo {author} {\bibfnamefont {A.}~\bibnamefont
  {Marinoni}}, \bibinfo {author} {\bibfnamefont {G.~R.}\ \bibnamefont {McKee}},
  \bibinfo {author} {\bibfnamefont {C.~C.}\ \bibnamefont {Petty}}, \bibinfo
  {author} {\bibfnamefont {J.~C.}\ \bibnamefont {Rost}}, \bibinfo {author}
  {\bibfnamefont {L.}~\bibnamefont {Schmitz}}, \bibinfo {author} {\bibfnamefont
  {G.}~\bibnamefont {Wang}}, \bibinfo {author} {\bibfnamefont {S.}~\bibnamefont
  {Zemedkun}}, \ and\ \bibinfo {author} {\bibfnamefont {L.}~\bibnamefont
  {Zeng}},\ }\href {\doibase 10.1063/1.4948723} {\bibfield  {journal} {\bibinfo
   {journal} {Phys. Plasmas}\ }\textbf {\bibinfo {volume} {23}},\ \bibinfo
  {pages} {056112} (\bibinfo {year} {2016})}\BibitemShut {NoStop}%
\bibitem [{\citenamefont {Howard}\ \emph {et~al.}(2016)\citenamefont {Howard},
  \citenamefont {Holland}, \citenamefont {White}, \citenamefont {Greenwald},
  \citenamefont {Candy},\ and\ \citenamefont {Creely}}]{howard:2016}%
  \BibitemOpen
  \bibfield  {author} {\bibinfo {author} {\bibfnamefont {N.~T.}\ \bibnamefont
  {Howard}}, \bibinfo {author} {\bibfnamefont {C.}~\bibnamefont {Holland}},
  \bibinfo {author} {\bibfnamefont {A.~E.}\ \bibnamefont {White}}, \bibinfo
  {author} {\bibfnamefont {M.}~\bibnamefont {Greenwald}}, \bibinfo {author}
  {\bibfnamefont {J.}~\bibnamefont {Candy}}, \ and\ \bibinfo {author}
  {\bibfnamefont {A.~J.}\ \bibnamefont {Creely}},\ }\href {\doibase
  10.1063/1.4946028} {\bibfield  {journal} {\bibinfo  {journal} {Phys.
  Plasmas}\ }\textbf {\bibinfo {volume} {23}},\ \bibinfo {pages} {056109}
  (\bibinfo {year} {2016})}\BibitemShut {NoStop}%
\bibitem [{\citenamefont {Idomura}(2014)}]{idomura:2014}%
  \BibitemOpen
  \bibfield  {author} {\bibinfo {author} {\bibfnamefont {Y.}~\bibnamefont
  {Idomura}},\ }\href {\doibase 10.1063/1.4867180} {\bibfield  {journal}
  {\bibinfo  {journal} {Phys. Plasmas}\ }\textbf {\bibinfo {volume} {21}},\
  \bibinfo {pages} {022517} (\bibinfo {year} {2014})}\BibitemShut {NoStop}%
\bibitem [{\citenamefont {Idomura}\ and\ \citenamefont
  {Nakata}(2014)}]{idomura:2014b}%
  \BibitemOpen
  \bibfield  {author} {\bibinfo {author} {\bibfnamefont {Y.}~\bibnamefont
  {Idomura}}\ and\ \bibinfo {author} {\bibfnamefont {M.}~\bibnamefont
  {Nakata}},\ }\href {\doibase 10.1063/1.4867379} {\bibfield  {journal}
  {\bibinfo  {journal} {Phys. Plasmas}\ }\textbf {\bibinfo {volume} {21}},\
  \bibinfo {pages} {020706} (\bibinfo {year} {2014})}\BibitemShut {NoStop}%
\bibitem [{\citenamefont {Dif-Pradalier}\ \emph {et~al.}(2015)\citenamefont
  {Dif-Pradalier}, \citenamefont {Hornung}, \citenamefont {Ghendrih},
  \citenamefont {Sarazin}, \citenamefont {Clairet}, \citenamefont {Vermare},
  \citenamefont {Diamond}, \citenamefont {Abiteboul}, \citenamefont
  {Cartier-Michaud}, \citenamefont {Ehrlacher}, \citenamefont {Est\`eve},
  \citenamefont {Garbet}, \citenamefont {Grandgirard}, \citenamefont
  {G\"urcan}, \citenamefont {Hennequin}, \citenamefont {Kosuga}, \citenamefont
  {Latu}, \citenamefont {Maget}, \citenamefont {Morel}, \citenamefont
  {Norscini}, \citenamefont {Sabot},\ and\ \citenamefont
  {Storelli}}]{difpradalier:2015}%
  \BibitemOpen
  \bibfield  {author} {\bibinfo {author} {\bibfnamefont {G.}~\bibnamefont
  {Dif-Pradalier}}, \bibinfo {author} {\bibfnamefont {G.}~\bibnamefont
  {Hornung}}, \bibinfo {author} {\bibfnamefont {P.}~\bibnamefont {Ghendrih}},
  \bibinfo {author} {\bibfnamefont {Y.}~\bibnamefont {Sarazin}}, \bibinfo
  {author} {\bibfnamefont {F.}~\bibnamefont {Clairet}}, \bibinfo {author}
  {\bibfnamefont {L.}~\bibnamefont {Vermare}}, \bibinfo {author} {\bibfnamefont
  {P.~H.}\ \bibnamefont {Diamond}}, \bibinfo {author} {\bibfnamefont
  {J.}~\bibnamefont {Abiteboul}}, \bibinfo {author} {\bibfnamefont
  {T.}~\bibnamefont {Cartier-Michaud}}, \bibinfo {author} {\bibfnamefont
  {C.}~\bibnamefont {Ehrlacher}}, \bibinfo {author} {\bibfnamefont
  {D.}~\bibnamefont {Est\`eve}}, \bibinfo {author} {\bibfnamefont
  {X.}~\bibnamefont {Garbet}}, \bibinfo {author} {\bibfnamefont
  {V.}~\bibnamefont {Grandgirard}}, \bibinfo {author} {\bibfnamefont {O.~D.}\
  \bibnamefont {G\"urcan}}, \bibinfo {author} {\bibfnamefont {P.}~\bibnamefont
  {Hennequin}}, \bibinfo {author} {\bibfnamefont {Y.}~\bibnamefont {Kosuga}},
  \bibinfo {author} {\bibfnamefont {G.}~\bibnamefont {Latu}}, \bibinfo {author}
  {\bibfnamefont {P.}~\bibnamefont {Maget}}, \bibinfo {author} {\bibfnamefont
  {P.}~\bibnamefont {Morel}}, \bibinfo {author} {\bibfnamefont
  {C.}~\bibnamefont {Norscini}}, \bibinfo {author} {\bibfnamefont
  {R.}~\bibnamefont {Sabot}}, \ and\ \bibinfo {author} {\bibfnamefont
  {A.}~\bibnamefont {Storelli}},\ }\href {\doibase
  10.1103/PhysRevLett.114.085004} {\bibfield  {journal} {\bibinfo  {journal}
  {Phys. Rev. Lett.}\ }\textbf {\bibinfo {volume} {114}},\ \bibinfo {pages}
  {085004} (\bibinfo {year} {2015})}\BibitemShut {NoStop}%
\bibitem [{\citenamefont {Kim}\ \emph {et~al.}(2017{\natexlab{a}})\citenamefont
  {Kim}, \citenamefont {Chang}, \citenamefont {Seo}, \citenamefont {Ku},\ and\
  \citenamefont {Choe}}]{kim:2017a}%
  \BibitemOpen
  \bibfield  {author} {\bibinfo {author} {\bibfnamefont {K.}~\bibnamefont
  {Kim}}, \bibinfo {author} {\bibfnamefont {C.~S.}\ \bibnamefont {Chang}},
  \bibinfo {author} {\bibfnamefont {J.}~\bibnamefont {Seo}}, \bibinfo {author}
  {\bibfnamefont {S.}~\bibnamefont {Ku}}, \ and\ \bibinfo {author}
  {\bibfnamefont {W.}~\bibnamefont {Choe}},\ }\href {\doibase
  10.1063/1.4974777} {\bibfield  {journal} {\bibinfo  {journal} {Phys.
  Plasmas}\ }\textbf {\bibinfo {volume} {24}},\ \bibinfo {pages} {012306}
  (\bibinfo {year} {2017}{\natexlab{a}})}\BibitemShut {NoStop}%
\bibitem [{\citenamefont {Kim}\ \emph {et~al.}(2017{\natexlab{b}})\citenamefont
  {Kim}, \citenamefont {Kwon}, \citenamefont {Chang}, \citenamefont {Seo},
  \citenamefont {Ku},\ and\ \citenamefont {Choe}}]{kim:2017b}%
  \BibitemOpen
  \bibfield  {author} {\bibinfo {author} {\bibfnamefont {K.}~\bibnamefont
  {Kim}}, \bibinfo {author} {\bibfnamefont {J.-M.}\ \bibnamefont {Kwon}},
  \bibinfo {author} {\bibfnamefont {C.~S.}\ \bibnamefont {Chang}}, \bibinfo
  {author} {\bibfnamefont {J.}~\bibnamefont {Seo}}, \bibinfo {author}
  {\bibfnamefont {S.}~\bibnamefont {Ku}}, \ and\ \bibinfo {author}
  {\bibfnamefont {W.}~\bibnamefont {Choe}},\ }\href {\doibase
  10.1063/1.4984991} {\bibfield  {journal} {\bibinfo  {journal} {Phys.
  Plasmas}\ }\textbf {\bibinfo {volume} {24}},\ \bibinfo {pages} {062302}
  (\bibinfo {year} {2017}{\natexlab{b}})}\BibitemShut {NoStop}%
\bibitem [{\citenamefont {Waltz}(2005)}]{waltz:2005}%
  \BibitemOpen
  \bibfield  {author} {\bibinfo {author} {\bibfnamefont {R.~E.}\ \bibnamefont
  {Waltz}},\ }\href {\doibase 10.13182/FST05-A1059} {\bibfield  {journal}
  {\bibinfo  {journal} {Fusion Sci. Technol.}\ }\textbf {\bibinfo {volume}
  {48}},\ \bibinfo {pages} {1051} (\bibinfo {year} {2005})}\BibitemShut
  {NoStop}%
\bibitem [{\citenamefont {Shimada}\ \emph {et~al.}(2007)\citenamefont
  {Shimada}, \citenamefont {Campbell}, \citenamefont {Mukhovatov},
  \citenamefont {Fujiwara}, \citenamefont {Kirneva}, \citenamefont {Lackner},
  \citenamefont {Nagami}, \citenamefont {Pustovitov}, \citenamefont {Uckan},
  \citenamefont {Wesley}, \citenamefont {Asakura}, \citenamefont {Costley},
  \citenamefont {Donné}, \citenamefont {Doyle}, \citenamefont {Fasoli},
  \citenamefont {Gormezano}, \citenamefont {Gribov}, \citenamefont {Gruber},
  \citenamefont {Hender}, \citenamefont {Houlberg}, \citenamefont {Ide},
  \citenamefont {Kamada}, \citenamefont {Leonard}, \citenamefont {Lipschultz},
  \citenamefont {Loarte}, \citenamefont {Miyamoto}, \citenamefont {Mukhovatov},
  \citenamefont {Osborne}, \citenamefont {Polevoi},\ and\ \citenamefont
  {Sips}}]{shimada:2007}%
  \BibitemOpen
  \bibfield  {author} {\bibinfo {author} {\bibfnamefont {M.}~\bibnamefont
  {Shimada}}, \bibinfo {author} {\bibfnamefont {D.}~\bibnamefont {Campbell}},
  \bibinfo {author} {\bibfnamefont {V.}~\bibnamefont {Mukhovatov}}, \bibinfo
  {author} {\bibfnamefont {M.}~\bibnamefont {Fujiwara}}, \bibinfo {author}
  {\bibfnamefont {N.}~\bibnamefont {Kirneva}}, \bibinfo {author} {\bibfnamefont
  {K.}~\bibnamefont {Lackner}}, \bibinfo {author} {\bibfnamefont
  {M.}~\bibnamefont {Nagami}}, \bibinfo {author} {\bibfnamefont
  {V.}~\bibnamefont {Pustovitov}}, \bibinfo {author} {\bibfnamefont
  {N.}~\bibnamefont {Uckan}}, \bibinfo {author} {\bibfnamefont
  {J.}~\bibnamefont {Wesley}}, \bibinfo {author} {\bibfnamefont
  {N.}~\bibnamefont {Asakura}}, \bibinfo {author} {\bibfnamefont
  {A.}~\bibnamefont {Costley}}, \bibinfo {author} {\bibfnamefont
  {A.}~\bibnamefont {Donné}}, \bibinfo {author} {\bibfnamefont
  {E.}~\bibnamefont {Doyle}}, \bibinfo {author} {\bibfnamefont
  {A.}~\bibnamefont {Fasoli}}, \bibinfo {author} {\bibfnamefont
  {C.}~\bibnamefont {Gormezano}}, \bibinfo {author} {\bibfnamefont
  {Y.}~\bibnamefont {Gribov}}, \bibinfo {author} {\bibfnamefont
  {O.}~\bibnamefont {Gruber}}, \bibinfo {author} {\bibfnamefont
  {T.}~\bibnamefont {Hender}}, \bibinfo {author} {\bibfnamefont
  {W.}~\bibnamefont {Houlberg}}, \bibinfo {author} {\bibfnamefont
  {S.}~\bibnamefont {Ide}}, \bibinfo {author} {\bibfnamefont {Y.}~\bibnamefont
  {Kamada}}, \bibinfo {author} {\bibfnamefont {A.}~\bibnamefont {Leonard}},
  \bibinfo {author} {\bibfnamefont {B.}~\bibnamefont {Lipschultz}}, \bibinfo
  {author} {\bibfnamefont {A.}~\bibnamefont {Loarte}}, \bibinfo {author}
  {\bibfnamefont {K.}~\bibnamefont {Miyamoto}}, \bibinfo {author}
  {\bibfnamefont {V.}~\bibnamefont {Mukhovatov}}, \bibinfo {author}
  {\bibfnamefont {T.}~\bibnamefont {Osborne}}, \bibinfo {author} {\bibfnamefont
  {A.}~\bibnamefont {Polevoi}}, \ and\ \bibinfo {author} {\bibfnamefont
  {A.}~\bibnamefont {Sips}},\ }\href
  {http://stacks.iop.org/0029-5515/47/i=6/a=S01} {\bibfield  {journal}
  {\bibinfo  {journal} {Nucl. Fusion}\ }\textbf {\bibinfo {volume} {47}},\
  \bibinfo {pages} {S1} (\bibinfo {year} {2007})}\BibitemShut {NoStop}%
\bibitem [{\citenamefont {Candy}\ \emph {et~al.}(2009)\citenamefont {Candy},
  \citenamefont {Holland}, \citenamefont {Waltz}, \citenamefont {Fahey},\ and\
  \citenamefont {Belli}}]{candy:2009}%
  \BibitemOpen
  \bibfield  {author} {\bibinfo {author} {\bibfnamefont {J.}~\bibnamefont
  {Candy}}, \bibinfo {author} {\bibfnamefont {C.}~\bibnamefont {Holland}},
  \bibinfo {author} {\bibfnamefont {R.~E.}\ \bibnamefont {Waltz}}, \bibinfo
  {author} {\bibfnamefont {M.~R.}\ \bibnamefont {Fahey}}, \ and\ \bibinfo
  {author} {\bibfnamefont {E.}~\bibnamefont {Belli}},\ }\href {\doibase
  10.1063/1.3167820} {\bibfield  {journal} {\bibinfo  {journal} {Phys.
  Plasmas}\ }\textbf {\bibinfo {volume} {16}},\ \bibinfo {pages} {060704}
  (\bibinfo {year} {2009})}\BibitemShut {NoStop}%
\bibitem [{\citenamefont {Barnes}\ \emph {et~al.}(2010)\citenamefont {Barnes},
  \citenamefont {Abel}, \citenamefont {Dorland}, \citenamefont {G\"{o}rler},
  \citenamefont {Hammett},\ and\ \citenamefont {Jenko}}]{barnes:2010}%
  \BibitemOpen
  \bibfield  {author} {\bibinfo {author} {\bibfnamefont {M.}~\bibnamefont
  {Barnes}}, \bibinfo {author} {\bibfnamefont {I.~G.}\ \bibnamefont {Abel}},
  \bibinfo {author} {\bibfnamefont {W.}~\bibnamefont {Dorland}}, \bibinfo
  {author} {\bibfnamefont {T.}~\bibnamefont {G\"{o}rler}}, \bibinfo {author}
  {\bibfnamefont {G.~W.}\ \bibnamefont {Hammett}}, \ and\ \bibinfo {author}
  {\bibfnamefont {F.}~\bibnamefont {Jenko}},\ }\href {\doibase
  10.1063/1.3323082} {\bibfield  {journal} {\bibinfo  {journal} {Phys.
  Plasmas}\ }\textbf {\bibinfo {volume} {17}},\ \bibinfo {pages} {056109}
  (\bibinfo {year} {2010})}\BibitemShut {NoStop}%
\bibitem [{\citenamefont {Ida}\ \emph {et~al.}(2015)\citenamefont {Ida},
  \citenamefont {Shi}, \citenamefont {Sun}, \citenamefont {Inagaki},
  \citenamefont {Kamiya}, \citenamefont {Rice}, \citenamefont {Tamura},
  \citenamefont {Diamond}, \citenamefont {Dif-Pradalier}, \citenamefont {Zou},
  \citenamefont {Itoh}, \citenamefont {Sugita}, \citenamefont {Gürcan},
  \citenamefont {Estrada}, \citenamefont {Hidalgo}, \citenamefont {Hahm},
  \citenamefont {Field}, \citenamefont {Ding}, \citenamefont {Sakamoto},
  \citenamefont {Oldenbürger}, \citenamefont {Yoshinuma}, \citenamefont
  {Kobayashi}, \citenamefont {Jiang}, \citenamefont {Hahn}, \citenamefont
  {Jeon}, \citenamefont {Hong}, \citenamefont {Kosuga}, \citenamefont {Dong},\
  and\ \citenamefont {Itoh}}]{ida:2015}%
  \BibitemOpen
  \bibfield  {author} {\bibinfo {author} {\bibfnamefont {K.}~\bibnamefont
  {Ida}}, \bibinfo {author} {\bibfnamefont {Z.}~\bibnamefont {Shi}}, \bibinfo
  {author} {\bibfnamefont {H.}~\bibnamefont {Sun}}, \bibinfo {author}
  {\bibfnamefont {S.}~\bibnamefont {Inagaki}}, \bibinfo {author} {\bibfnamefont
  {K.}~\bibnamefont {Kamiya}}, \bibinfo {author} {\bibfnamefont
  {J.}~\bibnamefont {Rice}}, \bibinfo {author} {\bibfnamefont {N.}~\bibnamefont
  {Tamura}}, \bibinfo {author} {\bibfnamefont {P.}~\bibnamefont {Diamond}},
  \bibinfo {author} {\bibfnamefont {G.}~\bibnamefont {Dif-Pradalier}}, \bibinfo
  {author} {\bibfnamefont {X.}~\bibnamefont {Zou}}, \bibinfo {author}
  {\bibfnamefont {K.}~\bibnamefont {Itoh}}, \bibinfo {author} {\bibfnamefont
  {S.}~\bibnamefont {Sugita}}, \bibinfo {author} {\bibfnamefont
  {O.}~\bibnamefont {Gürcan}}, \bibinfo {author} {\bibfnamefont
  {T.}~\bibnamefont {Estrada}}, \bibinfo {author} {\bibfnamefont
  {C.}~\bibnamefont {Hidalgo}}, \bibinfo {author} {\bibfnamefont
  {T.}~\bibnamefont {Hahm}}, \bibinfo {author} {\bibfnamefont {A.}~\bibnamefont
  {Field}}, \bibinfo {author} {\bibfnamefont {X.}~\bibnamefont {Ding}},
  \bibinfo {author} {\bibfnamefont {Y.}~\bibnamefont {Sakamoto}}, \bibinfo
  {author} {\bibfnamefont {S.}~\bibnamefont {Oldenbürger}}, \bibinfo {author}
  {\bibfnamefont {M.}~\bibnamefont {Yoshinuma}}, \bibinfo {author}
  {\bibfnamefont {T.}~\bibnamefont {Kobayashi}}, \bibinfo {author}
  {\bibfnamefont {M.}~\bibnamefont {Jiang}}, \bibinfo {author} {\bibfnamefont
  {S.}~\bibnamefont {Hahn}}, \bibinfo {author} {\bibfnamefont {Y.}~\bibnamefont
  {Jeon}}, \bibinfo {author} {\bibfnamefont {S.}~\bibnamefont {Hong}}, \bibinfo
  {author} {\bibfnamefont {Y.}~\bibnamefont {Kosuga}}, \bibinfo {author}
  {\bibfnamefont {J.}~\bibnamefont {Dong}}, \ and\ \bibinfo {author}
  {\bibfnamefont {S.-I.}\ \bibnamefont {Itoh}},\ }\href
  {http://stacks.iop.org/0029-5515/55/i=1/a=013022} {\bibfield  {journal}
  {\bibinfo  {journal} {Nucl. Fusion}\ }\textbf {\bibinfo {volume} {55}},\
  \bibinfo {pages} {013022} (\bibinfo {year} {2015})}\BibitemShut {NoStop}%
\bibitem [{\citenamefont {Waltz}\ \emph {et~al.}(2005)\citenamefont {Waltz},
  \citenamefont {Candy}, \citenamefont {Hinton}, \citenamefont {Estrada-Mila},\
  and\ \citenamefont {Kinsey}}]{waltz:2005b}%
  \BibitemOpen
  \bibfield  {author} {\bibinfo {author} {\bibfnamefont {R.}~\bibnamefont
  {Waltz}}, \bibinfo {author} {\bibfnamefont {J.}~\bibnamefont {Candy}},
  \bibinfo {author} {\bibfnamefont {F.}~\bibnamefont {Hinton}}, \bibinfo
  {author} {\bibfnamefont {C.}~\bibnamefont {Estrada-Mila}}, \ and\ \bibinfo
  {author} {\bibfnamefont {J.}~\bibnamefont {Kinsey}},\ }\href
  {http://stacks.iop.org/0029-5515/45/i=7/a=023} {\bibfield  {journal}
  {\bibinfo  {journal} {Nucl. Fusion}\ }\textbf {\bibinfo {volume} {45}},\
  \bibinfo {pages} {741} (\bibinfo {year} {2005})}\BibitemShut {NoStop}%
\bibitem [{\citenamefont {Waltz}\ \emph {et~al.}(2011)\citenamefont {Waltz},
  \citenamefont {Staebler},\ and\ \citenamefont {Solomon}}]{waltz:2011}%
  \BibitemOpen
  \bibfield  {author} {\bibinfo {author} {\bibfnamefont {R.~E.}\ \bibnamefont
  {Waltz}}, \bibinfo {author} {\bibfnamefont {G.~M.}\ \bibnamefont {Staebler}},
  \ and\ \bibinfo {author} {\bibfnamefont {W.~M.}\ \bibnamefont {Solomon}},\
  }\href {\doibase 10.1063/1.3579481} {\bibfield  {journal} {\bibinfo
  {journal} {Phys. Plasmas}\ }\textbf {\bibinfo {volume} {18}},\ \bibinfo
  {pages} {042504} (\bibinfo {year} {2011})}\BibitemShut {NoStop}%
\bibitem [{\citenamefont {Sugama}\ and\ \citenamefont
  {Horton}(1997)}]{sugama:1997}%
  \BibitemOpen
  \bibfield  {author} {\bibinfo {author} {\bibfnamefont {H.}~\bibnamefont
  {Sugama}}\ and\ \bibinfo {author} {\bibfnamefont {W.}~\bibnamefont
  {Horton}},\ }\href {\doibase 10.1063/1.872099} {\bibfield  {journal}
  {\bibinfo  {journal} {Phys. Plasmas}\ }\textbf {\bibinfo {volume} {4}},\
  \bibinfo {pages} {405} (\bibinfo {year} {1997})}\BibitemShut {NoStop}%
\bibitem [{\citenamefont {Sugama}\ and\ \citenamefont
  {Horton}(1998)}]{sugama:1998}%
  \BibitemOpen
  \bibfield  {author} {\bibinfo {author} {\bibfnamefont {H.}~\bibnamefont
  {Sugama}}\ and\ \bibinfo {author} {\bibfnamefont {W.}~\bibnamefont
  {Horton}},\ }\href {\doibase 10.1063/1.872941} {\bibfield  {journal}
  {\bibinfo  {journal} {Phys. Plasmas}\ }\textbf {\bibinfo {volume} {5}},\
  \bibinfo {pages} {2560} (\bibinfo {year} {1998})}\BibitemShut {NoStop}%
\bibitem [{\citenamefont {Abel}\ \emph {et~al.}(2013)\citenamefont {Abel},
  \citenamefont {Plunk}, \citenamefont {Wang}, \citenamefont {Barnes},
  \citenamefont {Cowley}, \citenamefont {Dorland},\ and\ \citenamefont
  {Schekochihin}}]{abel:2013}%
  \BibitemOpen
  \bibfield  {author} {\bibinfo {author} {\bibfnamefont {I.~G.}\ \bibnamefont
  {Abel}}, \bibinfo {author} {\bibfnamefont {G.~G.}\ \bibnamefont {Plunk}},
  \bibinfo {author} {\bibfnamefont {E.}~\bibnamefont {Wang}}, \bibinfo {author}
  {\bibfnamefont {M.}~\bibnamefont {Barnes}}, \bibinfo {author} {\bibfnamefont
  {S.~C.}\ \bibnamefont {Cowley}}, \bibinfo {author} {\bibfnamefont
  {W.}~\bibnamefont {Dorland}}, \ and\ \bibinfo {author} {\bibfnamefont
  {A.~A.}\ \bibnamefont {Schekochihin}},\ }\href
  {http://stacks.iop.org/0034-4885/76/i=11/a=116201} {\bibfield  {journal}
  {\bibinfo  {journal} {Rep. Prog. Phys.}\ }\textbf {\bibinfo {volume} {76}},\
  \bibinfo {pages} {116201} (\bibinfo {year} {2013})}\BibitemShut {NoStop}%
\bibitem [{\citenamefont {Crotinger}\ \emph {et~al.}(1997)\citenamefont
  {Crotinger}, \citenamefont {LoDestro}, \citenamefont {Pearlstein},
  \citenamefont {Tarditi}, \citenamefont {Casper},\ and\ \citenamefont
  {Hooper}}]{crotinger:1997}%
  \BibitemOpen
  \bibfield  {author} {\bibinfo {author} {\bibfnamefont {J.}~\bibnamefont
  {Crotinger}}, \bibinfo {author} {\bibfnamefont {L.}~\bibnamefont {LoDestro}},
  \bibinfo {author} {\bibfnamefont {L.}~\bibnamefont {Pearlstein}}, \bibinfo
  {author} {\bibfnamefont {A.}~\bibnamefont {Tarditi}}, \bibinfo {author}
  {\bibfnamefont {T.}~\bibnamefont {Casper}}, \ and\ \bibinfo {author}
  {\bibfnamefont {E.}~\bibnamefont {Hooper}},\ }\href {\doibase 10.2172/522508}
  {\emph {\bibinfo {title} {CORSICA: A comprehensive simulation of toroidal
  magnetic-fusion devices. Final report to the LDRD Program}}}\ (\bibinfo
  {year} {1997})\BibitemShut {NoStop}%
\bibitem [{\citenamefont {Shestakov}\ \emph {et~al.}(2003)\citenamefont
  {Shestakov}, \citenamefont {Cohen}, \citenamefont {Crotinger}, \citenamefont
  {LoDestro}, \citenamefont {Tarditi},\ and\ \citenamefont
  {Xu}}]{shestakov:2003}%
  \BibitemOpen
  \bibfield  {author} {\bibinfo {author} {\bibfnamefont {A.}~\bibnamefont
  {Shestakov}}, \bibinfo {author} {\bibfnamefont {R.}~\bibnamefont {Cohen}},
  \bibinfo {author} {\bibfnamefont {J.}~\bibnamefont {Crotinger}}, \bibinfo
  {author} {\bibfnamefont {L.}~\bibnamefont {LoDestro}}, \bibinfo {author}
  {\bibfnamefont {A.}~\bibnamefont {Tarditi}}, \ and\ \bibinfo {author}
  {\bibfnamefont {X.}~\bibnamefont {Xu}},\ }\href {\doibase
  http://dx.doi.org/10.1016/S0021-9991(02)00063-3} {\bibfield  {journal}
  {\bibinfo  {journal} {J. Comput. Phys.}\ }\textbf {\bibinfo {volume} {185}},\
  \bibinfo {pages} {399 } (\bibinfo {year} {2003})}\BibitemShut {NoStop}%
\bibitem [{tan()}]{tangocode}%
  \BibitemOpen
  \href {https://github.com/LLNL/tango} {}\Eprint
  {http://arxiv.org/abs/https://github.com/LLNL/tango}
  {https://github.com/LLNL/tango} \BibitemShut {NoStop}%
\bibitem [{\citenamefont {G\"{o}rler}\ \emph
  {et~al.}(2011{\natexlab{b}})\citenamefont {G\"{o}rler}, \citenamefont
  {Lapillonne}, \citenamefont {Brunner}, \citenamefont {Dannert}, \citenamefont
  {Jenko}, \citenamefont {Merz},\ and\ \citenamefont {Told}}]{gorler:2011}%
  \BibitemOpen
  \bibfield  {author} {\bibinfo {author} {\bibfnamefont {T.}~\bibnamefont
  {G\"{o}rler}}, \bibinfo {author} {\bibfnamefont {X.}~\bibnamefont
  {Lapillonne}}, \bibinfo {author} {\bibfnamefont {S.}~\bibnamefont {Brunner}},
  \bibinfo {author} {\bibfnamefont {T.}~\bibnamefont {Dannert}}, \bibinfo
  {author} {\bibfnamefont {F.}~\bibnamefont {Jenko}}, \bibinfo {author}
  {\bibfnamefont {F.}~\bibnamefont {Merz}}, \ and\ \bibinfo {author}
  {\bibfnamefont {D.}~\bibnamefont {Told}},\ }\href {\doibase
  http://dx.doi.org/10.1016/j.jcp.2011.05.034} {\bibfield  {journal} {\bibinfo
  {journal} {J. Comput. Phys.}\ }\textbf {\bibinfo {volume} {230}},\ \bibinfo
  {pages} {7053 } (\bibinfo {year} {2011}{\natexlab{b}})}\BibitemShut {NoStop}%
\bibitem [{gen()}]{genecode}%
  \BibitemOpen
  \href {www.genecode.org} {}\Eprint {http://arxiv.org/abs/www.genecode.org}
  {www.genecode.org} \BibitemShut {NoStop}%
\bibitem [{\citenamefont {Lapillonne}\ \emph {et~al.}(2009)\citenamefont
  {Lapillonne}, \citenamefont {Brunner}, \citenamefont {Dannert}, \citenamefont
  {Jolliet}, \citenamefont {Marinoni}, \citenamefont {Villard}, \citenamefont
  {G\"{o}rler}, \citenamefont {Jenko},\ and\ \citenamefont
  {Merz}}]{lapillonne:2009}%
  \BibitemOpen
  \bibfield  {author} {\bibinfo {author} {\bibfnamefont {X.}~\bibnamefont
  {Lapillonne}}, \bibinfo {author} {\bibfnamefont {S.}~\bibnamefont {Brunner}},
  \bibinfo {author} {\bibfnamefont {T.}~\bibnamefont {Dannert}}, \bibinfo
  {author} {\bibfnamefont {S.}~\bibnamefont {Jolliet}}, \bibinfo {author}
  {\bibfnamefont {A.}~\bibnamefont {Marinoni}}, \bibinfo {author}
  {\bibfnamefont {L.}~\bibnamefont {Villard}}, \bibinfo {author} {\bibfnamefont
  {T.}~\bibnamefont {G\"{o}rler}}, \bibinfo {author} {\bibfnamefont
  {F.}~\bibnamefont {Jenko}}, \ and\ \bibinfo {author} {\bibfnamefont
  {F.}~\bibnamefont {Merz}},\ }\href {\doibase 10.1063/1.3096710} {\bibfield
  {journal} {\bibinfo  {journal} {Phys. Plasmas}\ }\textbf {\bibinfo {volume}
  {16}},\ \bibinfo {pages} {032308} (\bibinfo {year} {2009})}\BibitemShut
  {NoStop}%
\bibitem [{\citenamefont {Candy}\ \emph {et~al.}(2004)\citenamefont {Candy},
  \citenamefont {Waltz},\ and\ \citenamefont {Dorland}}]{candy:2004}%
  \BibitemOpen
  \bibfield  {author} {\bibinfo {author} {\bibfnamefont {J.}~\bibnamefont
  {Candy}}, \bibinfo {author} {\bibfnamefont {R.~E.}\ \bibnamefont {Waltz}}, \
  and\ \bibinfo {author} {\bibfnamefont {W.}~\bibnamefont {Dorland}},\ }\href
  {\doibase 10.1063/1.1695358} {\bibfield  {journal} {\bibinfo  {journal}
  {Phys. Plasmas}\ }\textbf {\bibinfo {volume} {11}},\ \bibinfo {pages} {L25}
  (\bibinfo {year} {2004})}\BibitemShut {NoStop}%
\end{thebibliography}%

\end{document}